\documentclass[prl,twocolumn,showpacs,amsmath,amssymb,floatfix]{revtex4}

\newcommand{\ZZ}{{\mathbf{Z}}}
\newcommand{\NN}{{\mathbf{N}}}

\usepackage{graphicx}
\usepackage{dcolumn}
\usepackage{bm}

\begin{document}

\title{Non-deterministic density classification with
diffusive probabilistic cellular automata}
\author{Henryk Fuk\'s}%
 \email{hfuks@brocku.ca}
\affiliation{%
Department of Mathematics, Brock University\\
St. Catharines, Ontario L2S 3A1, Canada
}%

\date{\today}

\begin{abstract}
We present a probabilistic cellular automaton (CA) with two absorbing
states which performs classification of binary strings in
a non-deterministic sense. In a system evolving under this CA rule,
empty sites become occupied with a probability
proportional to the number of occupied sites in the neighborhood,
while occupied sites become empty with a probability proportional to
the number of empty sites in the neighborhood. The probability
that all sites become eventually occupied is equal to the density
of occupied sites in the initial string.
\end{abstract}

\maketitle

\section{Introduction}
Cellular automata (CA) and other spatially-extended discrete dynamical
systems are often used as models of complex systems with large
number of locally interacting components. One of the primary
problems encountered in constructing such models is the inverse
problem: the question how to find a local CA rule which would exhibit
the desired global behavior.

As a typical representative of the inverse problem, the
so-called density classification task \cite{GKL78} has been
extensively studied in recent years. The CA performing this task
should converge to a fixed point of all 1's if the initial
configuration contains more 1's than 0's, and to a fixed point of
all 0's if the converse is true.  While it has been proved~\cite{LB95}
that the  two-state rule performing this task does not exist, solutions of
modified tasks are possible if one allows more than one CA rule
\cite{paper4}, modifies specifications for the final configuration
\cite{Sipper96}, or assumes different boundary condition
\cite{SipperCR98}. Approximate solutions have been studied in the
context of genetic algorithms in one \cite{MitchellCH94} and two
dimensions~\cite{MoralesCM01}.

In what follows, we will define a probabilistic CA
which solves the density classification problem in the stochastic
sense, meaning that the probability that all sites become
eventually occupied is equal to the density of occupied sites in
the initial string.

We will assume that the dynamics takes place on a one-dimensional
lattice with periodic boundary conditions. Let $s_i(k)$ denotes
the state of the lattice site $i$ at time $k$, where $i \in \ZZ$,
$k\in \NN$. All operations on spatial indices $i$ are assumed to be
modulo $L$, where $L$ is the length of the lattice. We will
further assume that $s_i(k)\in \{0,1\}$, and we will say that the
site $i$ is occupied (empty) at time $k$ if $s_i(k)=1$
($s_i(k)=0$).

The dynamics of the system can be described as follows: empty
sites become occupied with a probability proportional to the
number of occupied sites in the neighborhood, while occupied
sites become empty with a probability proportional to the number
of empty sites in the neighborhood,  with all lattice sites
updated simultaneously at each time step. To be more precise, let
us denote by $P(s_i(k+1))| s_{i-1}(k),s_i(k),s_{i+1}(k))$ the
probability that the site $s_i(k)$ with nearest neighbors
$s_{i-1}(k),s_{i+1}(k)$ changes its state to $s_i(k+1)$ in a
single time step. The following set of transition probabilities
defines the aforementioned CA rule:
\begin{alignat}{2} \label{transprob}
P(1|0,0,0)&=0    & \qquad  P(1|0,0,1)&=p     \nonumber\\
P(1|0,1,0)&=1-2p & \qquad  P(1|0,1,1)&=1-p   \nonumber\\
P(1|1,0,0)&=p    & \qquad  P(1|1,0,1)&=2p    \nonumber\\
P(1|1,1,0)&=1-p  & \qquad  P(1|1,1,1)&=1,
\end{alignat}
where $p\in (0,1/2]$ (the remaining eight transition probabilities
can be obtained using $P(0|a,b,c)=1-P(1|a,b,c)$ for
$a,b,c\in\{0,1\}$).
The probabilistic CA defined by (\ref{transprob}) can be defined
explicitly if we introduce a set of iid random variables
$\{X_i\}_{i=0}^L$ with probability distribution $P(X_i=1)=p$,
$P(X_i=0)=1-p$, and another set $\{Y_i\}_{i=0}^L$ with probability
distribution $P(Y_i=1)=2p$, $P(Y_i=0)=1-2p$. Dynamics of the
 rule (\ref{transprob}) can then be described as
\begin{eqnarray} \label{micro}
s_i(k+1)=X_i (1-s_{i-1}) (1-s_{i}) s_{i+1} \nonumber \\
+(1-Y_i) (1-s_{i-1}) s_{i} (1-s_{i+1}) \nonumber \\
+(1-X_i) (1-s_{i-1}) s_{i} s_{i+1} \nonumber \\
+X_i s_{i-1} (1-s_{i}) (1-s_{i+1}) \nonumber \\
+Y_i s_{i-1} (1-s_{i}) s_{i+1} \nonumber \\
+(1-X_i) s_{i-1} s_{i} (1-s_{i+1}) \nonumber \\
+s_{i-1} s_{i} s_{i+1}.
\end{eqnarray}
To make the above formula easier to read, we omitted the time
argument, denoting $s_i(k)$ by $s_i$. After simplification and
reordering of terms, we obtain
\begin{eqnarray} \label{micro1}
&&s_i(k+1)=s_{i}-s_{i} Y_{i}+X_{i} s_{i-1}+X_{i} s_{i+1}\\
&+ &(s_{i-1} s_{i}+s_{i} s_{i+1}
-2 s_{i-1} s_{i} s_{i+1}+s_{i-1} s_{i+1}) (Y_{i}-2 X_{i}).\nonumber
\end{eqnarray}
\section{Difference and differential equations}
The state of the system at the time $k$ is determined by the
states of all lattice sites and is described by the Boolean random
field $\mathbf{s}(k)=\{s_i(k): i=0\ldots L\}$. The Boolean field
$\{\mathbf{s}(k): k=0,1,2,\ldots\}$ is then a Markov stochastic
process. Denoting by $E_{\mathbf{s}(0)}$ the expectation of this
Markov process when the initial configuration is $\mathbf{s}(0)$
we will now define the expected local density of occupied sites by
$\rho_i(k)=E_{\mathbf{s}(0)}\left[ s_i(k) \right]$. The expected
global density will be defined as
\begin{equation} \label{global}
\rho(k)=L^{-1} \sum_{i=0}^L \rho_i(k).
\end{equation}
While both $\rho_i(k)$ and $\rho(k)$ depend on the initial
configuration $\mathbf{s}(0)$, we will drop this dependence to simplify
notation. We will assume that the initial configuration
is exactly known (non-stochastic), hence $\rho(0)=\sum_{i=0}^L s_i(0)$
is the fraction of initially occupied sites.

Taking expectation value of both sides of
(\ref{micro1}), and taking into account that
$E_{\mathbf{s}(0)}\left[ Y_i-2X_i \right]=0$,
we obtain the following difference equation
\begin{equation} \label{dismf}
\rho_i(k+1)=\rho_i(k)+p \left( \rho_{i+1}(k)+ \rho_{i-1}(k)-2
\rho_i(k)\right).
\end{equation}
After summing over all lattice sites this yields
\begin{equation} \label{constrho}
\rho(k+1)=\rho(k),
\end{equation}
which means that the expected global density should be constant,
independently of the value of parameter $p$ and independently of
the initial configuration $\mathbf{s}(0)$. We can therefore say that the
probabilistic CA defined in (\ref{micro}) is
analogous to \emph{conservative CA}, i.e.,  deterministic CA which
preserve the number of occupied sites
\cite{Hattori91,paper8,paper10,pivato}.

Note that up to now we have not made any approximations, i.e.,
both (\ref{dismf}) and (\ref{constrho}) are exact. We can,
however, consider limiting behaviour of ($\ref{dismf}$) when the
physical distance between lattice sites and the size of the time
step simultaneously go to zero, using a similar procedure as
described in \cite{Law2000}. Let $x=\epsilon i$ and $t= \tau k$.
Now in (\ref{dismf}) we can replace $\rho(i,k)$ by $\rho(x,t)$,
$\rho(i \pm 1,k)$ by $\rho(x \pm \epsilon, t)$ and $\rho(i,k+1)$
by $\rho(x,t+\tau)$, which results in the following equation:
\begin{equation*}
\rho(r,t+\tau)=\rho(x,t)+p \left( \rho(x+\epsilon,t)+
\rho(x-\epsilon,t)-2 \rho(x,t)\right).
\end{equation*}
We will consider diffusive scaling in which time scales as a
square of the spatial length, meaning that $\tau=\epsilon^2$.
Taking Taylor expansion of the above equation in powers of
$\epsilon$ up to the second order we obtain
\begin{equation} \label{difeq}
\partial_t \rho=p \partial_x^2 \rho,
\end{equation}
i.e., the standard diffusion equation. Due to the form
of~(\ref{difeq}), in what follows we will refer to the process
defined in~(\ref{micro}) as \emph{diffusive probabilistic cellular
automaton (DPCA).}

\section{Absorption probability}
We will now present some simulation results illustrating
dynamics of DPCA. Since main features of DPCA remain
qualitatively the same for all values of the parameter $p$
in the interval $(0,1/2]$, we have chosen $p=0.25$
as a representative value to perform all subsequent
simulations.

\begin{figure}
\begin{center}
\includegraphics[scale=0.635]{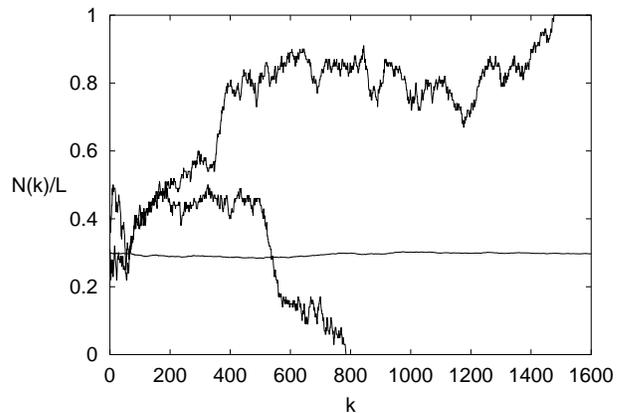}
\caption{\label{fig1} Fraction of occupied sites $N(k)/L$
as a function of time $k$ for two sample trajectories starting from
identical initial configuration with $N(0)=30$, $L=100$, $p=0.3$.
The third, almost horizontal line represents average of  $1000$
such trajectories.}
\end{center}
\end{figure}

\begin{figure}
\begin{center}
(a)\includegraphics[scale=0.635]{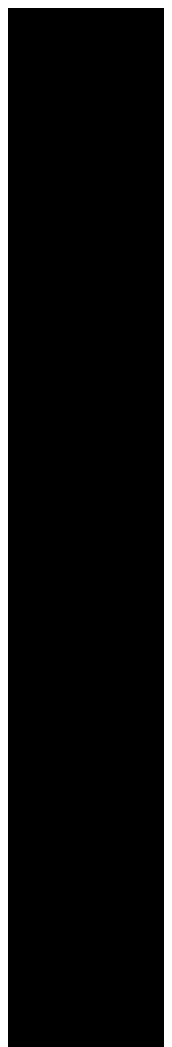}
\includegraphics[scale=0.635]{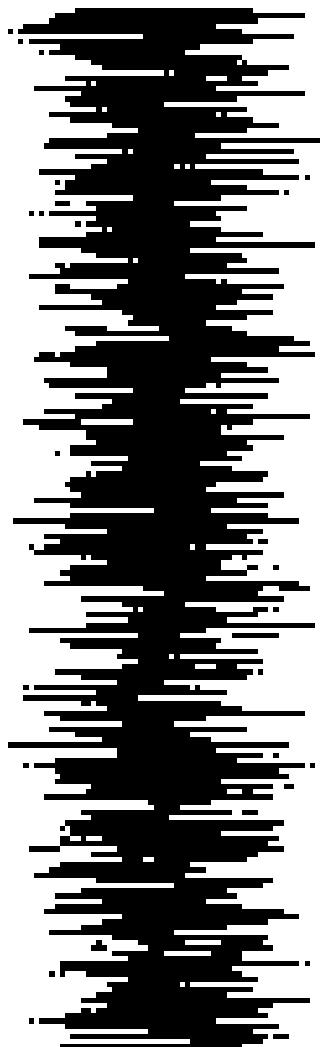}(b)\\
(c)\includegraphics[scale=0.635]{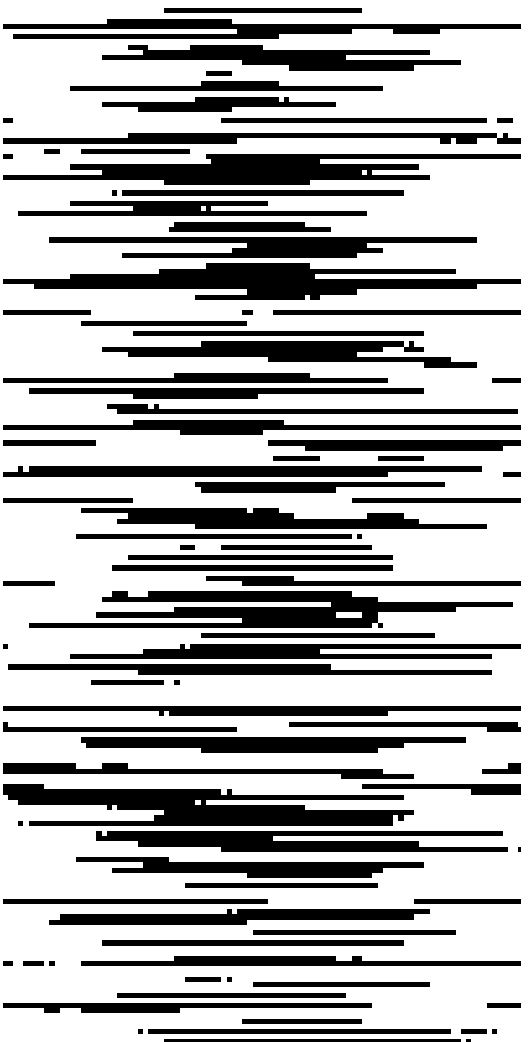}
\includegraphics[scale=0.635]{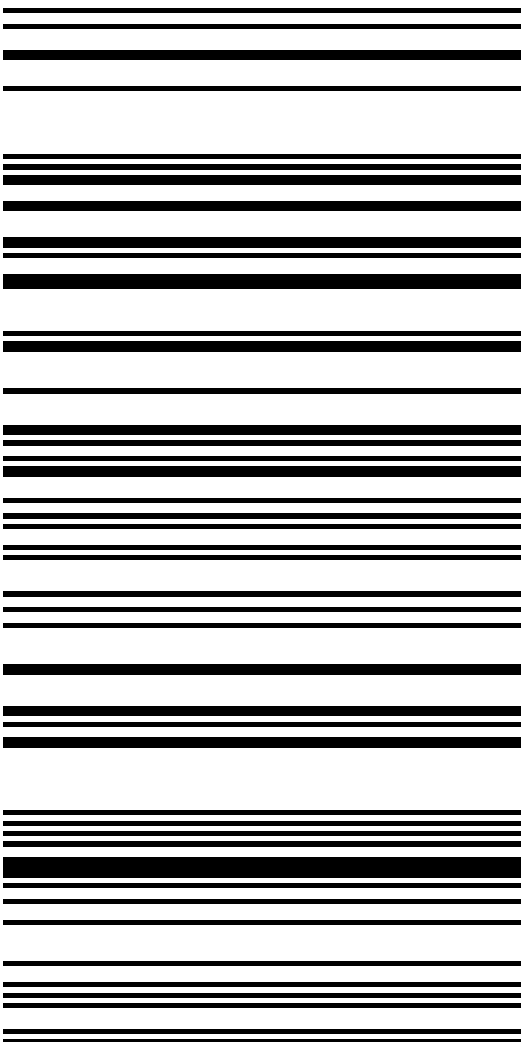}(d)
\caption{\label{fig2} Multiple realizations of
the CA evolution at (a) $k=0$, (b) $k=100$, (c) $k=1000$
and (d) $k=12000$. Each vertical line corresponds to different
realization of the process on $L=100$ lattice. Black pixels representing
occupied sites and white pixels empty sites. 200 different realizations
of the process are shown.}
\end{center}
\end{figure}

Let $N(k)=\sum_{i=1}^L s_i(k)$ be the number of occupied sites at
time $k$. If we start with $N(0)=0$, then $N(k)=0$ for all
$k>0$. Similarly, if $N(0)=L$, then $N(k)=L$ for all $k>0$. The
DPCA has thus two absorbing states, corresponding to all empty
sites
(to be referred to as $\mathbf{0}$) and to all occupied sites (to
be referred to as $\mathbf{1}$). If we start with $0<N(0)<L$,
then the graph of $N(k)$ resembles a random walk, as shown in
Figure~\ref{fig1}.
Both sample trajectories shown there eventually end in the
absorbing state, one of them in $\mathbf{0}$, another one in
$\mathbf{1}$. This is a general property of the DPCA:
regardless of the initial configuration, the system sooner or
later ends up in one of the two absorbing states. Although the
time required to reach the absorbing state can be large for a given
realization of the process, the expected value of the number of
time steps required to reach the absorbing state is finite, as it
is the case for all finite absorbing Markov chains \cite{kemeny}.
Figure~\ref{fig2} illustrates this property for
$L=100$  and the initial configuration with $30$
occupied sites clustered around the center, ie., located at $i=35,36,...,64$. All
other sites are empty. We start with an assembly of $200$ of such
initial configurations, all plotted as vertical lines
in which black pixels represent occupied sites, while
white pixels represent empty sites, as in Figure~\ref{fig2}a.
Each of these initial configurations evolves according to the
DPCA rule, and after $k=100$ ($k=1000$) iterations they are again
plotted as $200$ vertical lines in Figure~\ref{fig2}b (\ref{fig2}c).
After $12000$ iterations all $200$ configurations reach absorbing
states, as illustrated in Figure~\ref{fig2}d. Obviously, some
reach the state $\mathbf{1}$, while others $\mathbf{0}$, yet
it turns out that the fraction of configurations which ended
up in the state $\mathbf{1}$ is very close to $30\%$, the same
as the fraction of occupied sites at $k=0$.

To explain this phenomenon, let us define $u_N(k)$ to be the
probability that the number of occupied sites at time $k$ is $N$.
Since the Markov process $\{\mathbf{s}(k): k=0,1,2,\ldots\}$ is
finite and absorbing, no matter where the process starts, the
probability that after $k$ steps it is in an absorbing state tends
to $1$ as $k$ tends to infinity \cite{kemeny}. This implies that
\begin{eqnarray} \label{vanish}
 \lim_{k \rightarrow \infty}
 u_N(k)=0 \mbox{\ if $N\neq 0$ and $N \neq L$},\\
 \label{stay}
 \lim_{k \rightarrow \infty}\left( u_L(k)+ u_0(k)\right)=1.
\end{eqnarray}
The expected global density, as defined in (\ref{global}),
is independent of $k$, hence
\begin{equation}
\rho(0)=L^{-1} E_{\mathbf{s}(0)} \left[ N(k) \right]=L^{-1}
\sum_{N=1}^L N u_N(k).
\end{equation}
Taking the limit $k \rightarrow \infty$ of both sides of the above equation, and
using (\ref{vanish}) and (\ref{stay}), we obtain
\begin{eqnarray} \label{prob1}
 \lim_{k \rightarrow \infty} u_L(k)=\rho(0),\\
 \label{prob2}
  \lim_{k \rightarrow \infty} u_0(k)=1-\rho(0).
\end{eqnarray}
We have shown that \emph{the probability that the DPCA  reaches
the absorbing state $\mathbf{1}$ is equal to the initial fraction
of occupied sites $\rho(0)$. The probability that it reaches
$\mathbf{0}$ is $1-\rho(0)$.}
This is in agreement with the behavior observed in
Figure~\ref{fig2}.

The above can be viewed as a probabilistic generalization of the
density classification process.
In the standard (deterministic)
version of the density
classification problem we seek a CA rule which would
converge to $\mathbf{1}$ ($\mathbf{0}$) if the fraction of occupied
sites in the initial string is greater (less) than $1/2$, i.e.,
\begin{eqnarray} \label{det1}
\lim_{k \rightarrow \infty} u_L(k)=\Theta(\rho(0)),\\
\lim_{k \rightarrow \infty} u_0(k)=\Theta(1-\rho(0)),\label{det2}
\end{eqnarray}
where $\Theta(\cdot)$ is the step function defined
as $\Theta(x)=0$ if $x\leq1/2$, and $\Theta(x)=1$ if $x>1/2$.
Thus the difference between the deterministic and the probabilistic
density classification introduced here  is the replacement of the step
function  in (\ref{det1}-\ref{det2}) by the identity function
in (\ref{prob1}-\ref{prob2}).

As opposed to deterministically determined  outcome in the standard density
classification process, in DPCA it is just \emph{more probable} that the system
reaches  $\mathbf{1}$  then  $\mathbf{0}$ if the fraction of occupied sites in
the initial string is greater than $1/2$, and it is \emph{more probable} that it
reaches  $\mathbf{0}$  then $\mathbf{1}$ if the converse is true. Additionally,
DPCA can in some sense \emph{measure} concentration of occupied sites in the
initial string. If we want to know what is the initial density of occupied
sites, we need to run DPCA many times with the same initial condition until it
reaches the absorbing state, and observe how frequently it reaches $\mathbf{1}$.
This frequency will approximate $N(0)/L$, with accuracy increasing with the
number of experiments.

\section{Time to absorption}

Simulations shown in Figure~\ref{fig1} indicate that for large $k$
the system is typically in a state in which blocks of both empty
and occupied sites are relatively long. We can use this observation to obtain
the approximate dependence of the time required to reach the absorbing
state on the density of initial configuration.

If we assume that in a given configuration all occupied sites are
grouped in a few long continuous blocks, then the value of $N(k)$
cannot change too much in a single time step. For simplicity, let
us assume that the only allowed values of $\Delta
N(k)=N(k+1)-N(k)$ are $\{-2,-1,0,1,2\}$.
\begin{figure}
\begin{center}
\includegraphics[scale=0.635]{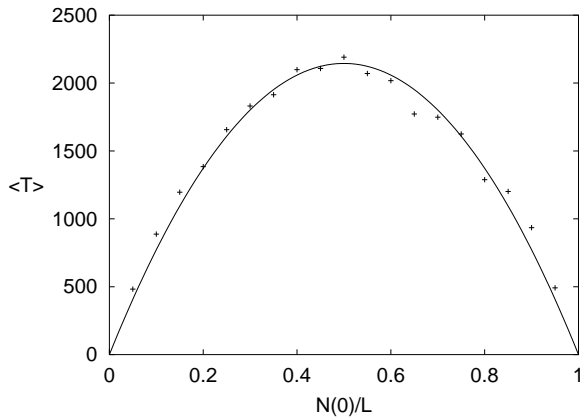}
\caption{\label{fig3} The average time required to reach the absorbing
state $\langle T \rangle$ as a function of the initial density
$\rho(0)=N(0)/L$  for $L=100$, $p=0.3$. Each data point ($+$)
represents the average taken over $1000$ realizations of the process
with identical initial configurations. The continuous line represents
fitted parabola $\langle T \rangle=\alpha \rho(0)(1-\rho(0))$.}
\end{center}
\end{figure}

Since $\rho(k)$ is time-independent, the probability that $N(k)$
increases by a given amount should be equal to the probability
that it decreases by the same amount in a single time step. In the
agreement with the above, let us define $p_j$ as the probability
that $\Delta N(k)$ takes the value $j$, so that $p_j$ is non-zero
for $j\in \{-2,-1,0,1,2\}$, where $p_{-2}=p_{2}$, $p_{-1}=p_{1}$,
and  $2p_2+2p_1+p_0=1$. If $T_z$ denotes the expected time to
reach the state $N(k)=0$ or $N(k)=L$ starting from $N(0)=z$, a
simple argument \cite{feller} yields the difference equation which
$T_z$ must satisfy
\begin{equation}
T_z=p_2 T_{z+2} + p_1 T_{z+1} + p_{0} T_{z} +p_1 T_{z-1} + p_2
T_{z-2} +1.
\end{equation}
The solution of this equation satisfying boundary conditions $T_0=0$
and $T_L=0$ is given by
\begin{equation} \label{atime}
T_z=\alpha \,  z(L-z)=\alpha \,  L^2\rho(0) (1-\rho(0)) ,
\end{equation}
where $\alpha=1/(8p_2+4p_1)$, meaning that the mean time
to absorption scales with lattice length as ${O} (L^2)$.
The above result
would remain valid even if we allowed further jumps than $\pm 2$
(although the form of the coefficient $\alpha$ would be
different).

In order to verify if this result holds for the DPCA, we performed
a series of numerical experiments, computing the average time
required to reach the absorbing state for 1000 realizations of the
DPCA process, for a range of initial densities. Results are shown
if Figure~\ref{fig3}. One can clearly see that data points are
aligned along a curve of parabolic shape, as expected from
(\ref{atime}).

\section{Conclusion}
The probabilistic CA introduced in this article
solves the density classification problem in a non-deterministic
sense. It is interesting to note that the DPCA conserves the
average number of occupied sites, similarly as deterministic rules
employed in  solutions of related  problems
mentioned in the introduction. Indeed, conservation of the
number of occupied sites  is a necessary condition for density
classification by CA if one allows modified output
configuration, as recently shown in \cite{CapcarrereS01}. This suggests
that a wider class of probabilistic CA conserving $\rho(k)$ might
be an useful paradigm in studying how locally interacting systems
compute global properties, and certainly deserves  further
attention.

\begin{center} \textbf{Acknowledgements} \end{center}
The author acknowledges financial support from the Natural
Sciences and Engineering Research Council of Canada.


\begin{thebibliography}{15}
\expandafter\ifx\csname natexlab\endcsname\relax\def\natexlab#1{#1}\fi
\expandafter\ifx\csname bibnamefont\endcsname\relax
  \def\bibnamefont#1{#1}\fi
\expandafter\ifx\csname bibfnamefont\endcsname\relax
  \def\bibfnamefont#1{#1}\fi
\expandafter\ifx\csname citenamefont\endcsname\relax
  \def\citenamefont#1{#1}\fi
\expandafter\ifx\csname url\endcsname\relax
  \def\url#1{\texttt{#1}}\fi
\expandafter\ifx\csname urlprefix\endcsname\relax\def\urlprefix{URL }\fi
\providecommand{\bibinfo}[2]{#2}
\providecommand{\eprint}[2][]{\url{#2}}

\bibitem[{\citenamefont{Gacs et~al.}(1987)\citenamefont{Gacs, Kurdymov, and
  Levin}}]{GKL78}
\bibinfo{author}{\bibfnamefont{P.}~\bibnamefont{Gacs}},
  \bibinfo{author}{\bibfnamefont{G.~L.} \bibnamefont{Kurdymov}},
  \bibnamefont{and} \bibinfo{author}{\bibfnamefont{L.~A.} \bibnamefont{Levin}},
  \bibinfo{journal}{Probl. Peredachi Inform.} \textbf{\bibinfo{volume}{14}},
  \bibinfo{pages}{92} (\bibinfo{year}{1987}).

\bibitem[{\citenamefont{Land and Belew}(1995)}]{LB95}
\bibinfo{author}{\bibfnamefont{M.}~\bibnamefont{Land}} \bibnamefont{and}
  \bibinfo{author}{\bibfnamefont{R.~K.} \bibnamefont{Belew}},
  \bibinfo{journal}{Phys. Rev. Lett.} \textbf{\bibinfo{volume}{74}},
  \bibinfo{pages}{5148} (\bibinfo{year}{1995}).

\bibitem[{\citenamefont{Fuk{\'s}}(1997)}]{paper4}
\bibinfo{author}{\bibfnamefont{H.}~\bibnamefont{Fuk{\'s}}},
  \bibinfo{journal}{Phys. Rev. E} \textbf{\bibinfo{volume}{55}},
  \bibinfo{pages}{2081R} (\bibinfo{year}{1997}),
  \eprint{arXiv:comp-gas/9703001}.

\bibitem[{\citenamefont{Capcarr{\`e}re
  et~al.}(1996)\citenamefont{Capcarr{\`e}re, Sipper, and Tomassini}}]{Sipper96}
\bibinfo{author}{\bibfnamefont{M.~S.} \bibnamefont{Capcarr{\`e}re}},
  \bibinfo{author}{\bibfnamefont{M.}~\bibnamefont{Sipper}}, \bibnamefont{and}
  \bibinfo{author}{\bibfnamefont{M.}~\bibnamefont{Tomassini}},
  \bibinfo{journal}{Phys. Rev. Lett.} \textbf{\bibinfo{volume}{77}},
  \bibinfo{pages}{4969} (\bibinfo{year}{1996}).

\bibitem[{\citenamefont{Sipper et~al.}(1998)\citenamefont{Sipper,
  Capcarr{\`e}re, and Ronald}}]{SipperCR98}
\bibinfo{author}{\bibfnamefont{M.}~\bibnamefont{Sipper}},
  \bibinfo{author}{\bibfnamefont{M.~S.} \bibnamefont{Capcarr{\`e}re}},
  \bibnamefont{and} \bibinfo{author}{\bibfnamefont{E.}~\bibnamefont{Ronald}},
  \bibinfo{journal}{Int. J. Mod. Phys. C} \textbf{\bibinfo{volume}{9}},
  \bibinfo{pages}{899} (\bibinfo{year}{1998}).

\bibitem[{\citenamefont{Mitchell et~al.}(1994)\citenamefont{Mitchell,
  Crutchfield, and Hraber}}]{MitchellCH94}
\bibinfo{author}{\bibfnamefont{M.}~\bibnamefont{Mitchell}},
  \bibinfo{author}{\bibfnamefont{J.~P.} \bibnamefont{Crutchfield}},
  \bibnamefont{and} \bibinfo{author}{\bibfnamefont{P.~T.}
  \bibnamefont{Hraber}}, \bibinfo{journal}{Physica D}
  \textbf{\bibinfo{volume}{75}}, \bibinfo{pages}{361} (\bibinfo{year}{1994}).

\bibitem[{\citenamefont{Morales et~al.}(2001)\citenamefont{Morales,
  Crutchfield, and Mitchell}}]{MoralesCM01}
\bibinfo{author}{\bibfnamefont{F.~J.} \bibnamefont{Morales}},
  \bibinfo{author}{\bibfnamefont{J.~P.} \bibnamefont{Crutchfield}},
  \bibnamefont{and} \bibinfo{author}{\bibfnamefont{M.}~\bibnamefont{Mitchell}},
  \bibinfo{journal}{Parallel Comput.} \textbf{\bibinfo{volume}{27}},
  \bibinfo{pages}{571} (\bibinfo{year}{2001}).

\bibitem[{\citenamefont{Hattori and Takesue}(1991)}]{Hattori91}
\bibinfo{author}{\bibfnamefont{T.}~\bibnamefont{Hattori}} \bibnamefont{and}
  \bibinfo{author}{\bibfnamefont{S.}~\bibnamefont{Takesue}},
  \bibinfo{journal}{Physica D} \textbf{\bibinfo{volume}{49}},
  \bibinfo{pages}{295} (\bibinfo{year}{1991}).

\bibitem[{\citenamefont{Boccara and Fuk{\'s}}(1998)}]{paper8}
\bibinfo{author}{\bibfnamefont{N.}~\bibnamefont{Boccara}} \bibnamefont{and}
  \bibinfo{author}{\bibfnamefont{H.}~\bibnamefont{Fuk{\'s}}},
  \bibinfo{journal}{J. Phys. A: Math. Gen.} \textbf{\bibinfo{volume}{31}},
  \bibinfo{pages}{6007} (\bibinfo{year}{1998}),
  \eprint{arXiv:adap-org/9712003}.

\bibitem[{\citenamefont{Fuk{\'s}}(2000)}]{paper10}
\bibinfo{author}{\bibfnamefont{H.}~\bibnamefont{Fuk{\'s}}}, in
  \emph{\bibinfo{booktitle}{Hydrodynamic Limits and Related Topics,}}, edited
  by \bibinfo{editor}{\bibfnamefont{S.}~\bibnamefont{Feng}},
  \bibinfo{editor}{\bibfnamefont{A.~T.} \bibnamefont{Lawniczak}},
  \bibnamefont{and} \bibinfo{editor}{\bibfnamefont{R.~S.}
  \bibnamefont{Varadhan}} (\bibinfo{publisher}{AMS},
  \bibinfo{address}{Providence, RI}, \bibinfo{year}{2000}),
  \eprint{arXiv:nlin.CG/0207047}.

\bibitem[{\citenamefont{Pivato}(2001)}]{pivato}
\bibinfo{author}{\bibfnamefont{M.}~\bibnamefont{Pivato}}
  (\bibinfo{year}{2001}), \bibinfo{note}{preprint},
  \eprint{arXiv:math.DS/0111014}.

\bibitem[{\citenamefont{Lawniczak}(2000)}]{Law2000}
\bibinfo{author}{\bibfnamefont{A.}~\bibnamefont{Lawniczak}},
  \bibinfo{journal}{Transport Theory and Statistical Physics}
  \textbf{\bibinfo{volume}{29}}, \bibinfo{pages}{261} (\bibinfo{year}{2000}).

\bibitem[{\citenamefont{Kemeny and Snell}(1960)}]{kemeny}
\bibinfo{author}{\bibfnamefont{J.~G.} \bibnamefont{Kemeny}} \bibnamefont{and}
  \bibinfo{author}{\bibfnamefont{J.~L.} \bibnamefont{Snell}},
  \emph{\bibinfo{title}{Finite Markov Chains}} (\bibinfo{publisher}{D. Van
  Nostrand Co.}, \bibinfo{address}{Princeton, NJ}, \bibinfo{year}{1960}).

\bibitem[{\citenamefont{Feller}(1968)}]{feller}
\bibinfo{author}{\bibfnamefont{W.}~\bibnamefont{Feller}},
  \emph{\bibinfo{title}{An Introduction to Probability Theory and Its
  Applications}} (\bibinfo{publisher}{Wiley and Sons, Inc.},
  \bibinfo{address}{New York}, \bibinfo{year}{1968}).

\bibitem[{\citenamefont{Capcarr{\`e}re and Sipper}(2001)}]{CapcarrereS01}
\bibinfo{author}{\bibfnamefont{M.~S.} \bibnamefont{Capcarr{\`e}re}}
  \bibnamefont{and} \bibinfo{author}{\bibfnamefont{M.}~\bibnamefont{Sipper}},
  \bibinfo{journal}{Phys. Rev. E} \textbf{\bibinfo{volume}{64}},
  \bibinfo{pages}{036113} (\bibinfo{year}{2001}).

\end{thebibliography}

\end{document}